\documentclass[12pt,a4paper,final]{iopart}

\usepackage{iopams}  
\usepackage[breaklinks=true,colorlinks=true,linkcolor=blue,urlcolor=blue,citecolor=blue]{hyperref}

\usepackage[utf8]{inputenc}
\usepackage[T1]{fontenc}
\usepackage{mathptmx}

\usepackage[english]{babel}
\usepackage{hyperref}
\usepackage[pdftex]{graphicx}
\usepackage{amsthm}
\usepackage{xcolor}
\newcommand{\rev}[1]{{#1}}

\begin{document}

\title{Robinson–Trautman spacetimes in 
(2+1) dimensions}

\author{Alberto Saa}%
\address{Department of Applied Mathematics,  
 University of Campinas,  \\ 13083-859 Campinas, SP, Brazil.}
  \ead{asaa@ime.unicamp.br}

\begin{abstract}
We propose a Robinson-Trautman evolution in $(2+1)$-dimensional spacetime that
retains key structural features  of the four-dimensional case. We consider a recently studied exact family of metrics to select a nonstationary geometry with a cosmological constant, sourced by a null fluid. The metric is completely determined by a single positive function $P(u,\phi)$, while the corresponding matter content is encoded in a null-fluid density. Motivated by the role of the area-preserving Calabi flow in four dimensions, we introduce a fourth-order length-preserving evolution equation for $P(u,\phi)$ whose stationary configurations correspond, for negative cosmological constant, to boosted BTZ black holes. Numerical solutions strongly support the relaxation of generic regular initial data $P(0,\phi)$ toward the stationary sector. The resulting system provides a simple toy model for dissipative dynamics driven by null radiation in lower-dimensional gravity, with several structural similarities to phenomena associated with genuine gravitational radiation.
\end{abstract}


\submitto{\CQG}

\section{Introduction}

The Robinson-Trautman (RT) spacetime \cite{exact} has a distinguished place in General Relativity (GR).
It is perhaps
the simplest spacetime 
describing a compact source surrounded by gravitational waves. As an initial
value problem, the RT spacetime evolution is a well-posed problem, in the sense that any
regular initial data will evolve
smoothly according to the RT evolution equations towards a final state
corresponding to a remnant   Schwarzschild black-hole \cite{evolution}. The RT equations
have an interesting and elegant geometrical interpretation, they correspond to the two-dimensional area-preserving Calabi flow for the
  family  of $r=1$  two-dimensional spatial submanifold (unit spheres).
   The RT spacetime provides 
one of the cleanest laboratories for studying gravitational dissipation and final-state selection   as, for instance, 
  in  the problem of
gravitational wave recoil \cite{MacedoSaa}.  
Higher-dimensional generalizations preserve much of this structure \cite{D>4}. Although the algebraic classification
of the corresponding Weyl tensors 
 becomes more involved for $D>(3+1)$ spacetime dimensions, the essential picture remains the same: regular data evolve toward a higher-dimensional Schwarzschild remnant. The natural question is whether an analogue of this scenario survives in three spacetime dimensions, where the local degrees of freedom of the gravitational field are drastically reduced.

The answer is not obvious. In $(2+1)$ dimensions the 
Riemann tensor is completely determined by its traces (namely the Ricci tensor and the scalar curvature), any Ricci flat solution  will
be necessarily 
flat, implying the absence of standard gravitational waves for GR in three-dimensional spacetimes.
Such a behavior  is essentially unaltered   in the presence of a cosmological constant.
Nevertheless, the standard GR  with negative cosmological constant in 
 $(2+1)$ dimensions  
 does  admit 
a physically consistent black-hole solution, namely 
the BTZ black hole \cite{BTZ}, which indeed provides a rich   lower-dimensional black-hole geometry. This suggests looking for a lower dimensional RT-like setting in which the geometry would not be vacuum, but   sourced by a null radiating fluid,
and the evolution  would accommodate  a meaningful relaxation toward a BTZ-type remnant, mimicking 
physically and mathematically 
the higher dimensional
initial value problem.
\rev{Related pure-radiation configurations in three-dimensional gravity, including solutions with gyratons in massive gravity theories, have also been discussed recently in Ref.~\cite{Ercan}.}

The purpose of this work is precisely to formulate such a lower-dimensional analogue. Starting from the complete family of $(2+1)$-dimensional Robinson-Trautman solutions studied in detail in \cite{2+1}, we isolate a one-function sub-family  parametrized by $P(u,\phi)$, compute the induced null-fluid flux, and then propose an intrinsic fourth-order nonlinear evolution equation for $P$ that preserves the total length of the associated unit circle. We show that the stationary configurations form a normalized family of boosted BTZ states. We also study the linearized dynamics around the isotropic configuration, and discuss numerical evidence for the full nonlinear relaxation. 
It is important to   stress that our construction is not a lower-dimensional reduction of the four-dimensional vacuum problem; rather, it should be regarded as a dynamical toy model that captures many interesting features of the dynamics of real gravitational waves.

\section{The RT flow in (2+1) dimensions}

Recently, there has been significant activity in $(2+1)$ gravity, particularly  regarding  the subtle issues of asymptotic analysis. In addition to Ref. \cite{2+1}, which provides all conventions we adopt here, the main features of the Robinson-Trautman solutions in 
$(2 + 1)$ dimensions have been thoroughly examined in Refs. \cite{as1,as2,as3,as4}.
{
For the sake of completeness, we first revisit the field equations for the subfamily 
of solutions we will explore here. We employ the conventions of  \cite{2+1}, which include
\begin{equation}
G_{ab}+\Lambda g_{ab}=8\pi T_{ab}
\end{equation}
for the Einstein's equations. 
If the mass aspect is temporarily kept as an arbitrary function $\mu(u)$, the corresponding member of the $(2+1)$-dimensional Robinson--Trautman family can be written as (see \cite{2+1} for the details)
\begin{equation}
\label{metric}
ds^2 =   \left(\mu(u) + 2r(\ln P )_u +\Lambda r^2\right)du^2 - 2dudr + \frac{r^2}{P^2}d\phi^2,
\end{equation}
with $P=P(u,\phi)>0$. Direct substitution in   Einstein's equations gives $R=6\Lambda$, and all components of $G_{ab}+\Lambda g_{ab}$ vanish except
\begin{equation}
G_{uu}+\Lambda g_{uu}=\frac{1}{r}
\left[\mu(u)(\ln P)_u-\Delta(\ln P)_u-\frac{1}{2}\dot\mu(u)\right],
\end{equation}
where a dot denotes differentiation with respect to $u$, and
\begin{equation}
\Delta f=P\partial_\phi\left(P\partial_\phi f\right)
\end{equation}
can be seen as the one-dimensional Laplacian associated with the angular part of the metric. The corresponding source is therefore a  null fluid,
\begin{equation}
\label{Tuu}
T_{uu} = \frac{\mathcal{N}(u,\phi)}{r},  
\end{equation}
with
\begin{equation}
\label{N}
\mathcal{N}=\frac{1}{8\pi}
\left[\mu(u)(\ln P)_u-\Delta(\ln P)_u-\frac{1}{2}\dot\mu(u)\right].
\end{equation}

In the present paper we will restrict our attention to the constant-mass case $\mu(u)=m_0$.  
It corresponds clearly to a sub-family of the general solution (\ref{metric}) considered in detail in \cite{2+1,as2}. Furthermore,
  the underlying geometry becomes the standard BTZ metric when $P=1$
and $\Lambda <0$, while more general angular profiles $P(u,\phi)$ describe anisotropic null-radiation configurations within the exact Robinson-Trautman family of Ref. \cite{2+1}.
 The restriction $\mu(u)=m_0$ has a simple but important physical meaning. If the full mass aspect $\mu(u)$ is retained, equation (\ref{N}) contains the additional isotropic contribution $-\dot\mu/(16\pi)$. In particular, for the circularly symmetric case $P=1$, one obtains $\mathcal{N}=-\dot\mu/(16\pi)$, so that the usual Vaidya interpretation is recovered: a decreasing mass aspect, $\dot\mu<0$, corresponds to a positive outgoing null-fluid density. By contrast, the sector considered here freezes the BTZ mass parameter and removes this isotropic mass-loss channel. The null fluid in (\ref{Tuu}) should therefore be understood as an anisotropic null fluid distribution superposed on a fixed BTZ mass parameter, rather than as the radiation emitted by an ordinary Vaidya source losing mass. This is the sector in which the angular profile $P(u,\phi)$ alone controls both the geometry and the null fluid density.
}

A natural geometric quantity for the spacetime (\ref{metric}) associated with the $r=1$   section  is its total length,
\begin{equation}
\label{ell}
\ell(u) = \int_0^{2\pi} \frac{d\phi}{P(u,\phi)}  .
\end{equation}
Differentiating with respect to $u$ and using (\ref{N}) gives
\begin{equation}
\label{ell1}
\frac{d}{du}   \ell(u)  = -\frac{8\pi}{m_0}  \Phi_\mathcal{N},
\end{equation}
where the total  flux $\Phi_\mathcal{N} $  is given by
\begin{equation}
\label{ell2}
\Phi_\mathcal{N} = \int_0^{2\pi} \frac{\mathcal{N}(u,\phi)}{P(u,\phi)} d\phi.
\end{equation} 
{
For comparison, if the mass aspect $\mu(u)$ is not fixed, the same calculation gives
\begin{equation}
\frac{d}{du}   \ell(u) =-\frac{8\pi}{\mu(u)}\Phi_\mathcal{N}-\frac{\dot\mu(u)}{2\mu(u)}\ell(u).
\end{equation}
Equation (\ref{ell1}) is precisely the constant-mass specialization of this more general balance law.
}
As one can see, the evolution of $\ell(u)$ is entirely governed by the total   null-fluid flux. 
The  total length 
of the unit circle 
(\ref{ell}) will be constant during the evolution if and only if the total flux $\Phi_\mathcal{N} $ vanishes.
Notice that 
the second term in (\ref{N}) does not contribute to the total flux due to the ``one-dimensional Stokes theorem''.
 For the $D=(3+1)$ case, the equivalent of $\ell(u)$ is the area of the   unit sphere, which is indeed constant 
along the RT spacetime evolution. It is important to emphasize a key point here. Since $P(u,\phi)$ is assumed to be a
 positive function, for the total flux described by   (\ref{ell2}) to vanish, the density $ \mathcal{N}(u,\phi)$ cannot also be a strictly positive function. Instead, it must include regions on the circle with both positive and negative values. In other words, to ensure a constant $\ell(u)$, one must allow for violations of the null energy conditions associated with the energy-momentum tensor given by equation (\ref{Tuu}).
 
For $P=1$ and $\Lambda < 0$, the metric (\ref{metric}) reduces to the usual BTZ black hole, for which the
unit circle length is indeed constant $\ell(u)=2\pi$. 
Another stationary sector of direct relevance for us is obtained by taking  $P=P_0(\phi) = a + b\cos \phi + c\sin \phi$, with positive $a$ such that $a^2 > b^2 + c^2$. \rev{This inequality is exactly the condition ensuring $P_0(\phi)>0$ on the entire circle.} For this case,
we have
\begin{equation}
\ell = \frac{2\pi}{\sqrt{a^2 - b^2 - c^2}},
\end{equation}
and by imposing the normalization $\ell = 2\pi$ we have $a^2 - b^2 - c^2=1$, which can be parametrized as
\begin{equation}
\label{rocket}
P_0(\phi) =   \gamma\left(1 +   v\cos(\phi - \phi_0)\right),
\end{equation}
with $\gamma = \frac{1}{\sqrt{1-v^2}}$ and $ 0 \le v < 1$.
For this choice, the metric (\ref{metric}) corresponds  to a particular case of the Kinnersley photon-rocket spacetime discussed in Refs. \cite{2+1,r1,r2,r3}. In the present context, it is naturally interpreted as a
boosted BTZ black hole, {\em i.e.}, a
 BTZ black hole moving with constant speed $v$ in the spatial direction given by 
 $(\cos\phi_0,\sin\phi_0)$. Analogous situations also hold for higher dimensions, see \cite{r4}. The family (\ref{rocket}) will play the role of the stationary solutions of the $(2+1)$ RT flow we   introduced below.

For any smooth positive $P(u,\phi)$, the metric (\ref{metric}) is an exact solution of the Einstein equations with cosmological constant and the null-fluid source (\ref{Tuu}). We now introduce a distinguished evolution law for $P(u,\phi)$
specifically  designed to mimic the structural properties of the four-dimensional Robinson-Trautman equation. More precisely, we demand that:  
\begin{enumerate}
\item  the total length $\ell(u)$ given by (\ref{ell}) be preserved;
\item  regular initial data relax toward the stationary family (\ref{rocket});  
\item   the evolution be governed by a nonlinear fourth-order parabolic equation.
\end{enumerate} 
A natural candidate satisfying these requirements is
\begin{equation}
\label{rtf}
(\ln P )_u = -\kappa \Delta \left( P + \frac{1}{P}\Delta \ln P \right),
\end{equation}
where $\kappa>0$ is a constant setting the relaxation time scale. 
To understand why this is a natural candidate, we need to delve deeper into the third condition. The requirement for a nonlinear fourth-order parabolic equation is specifically aimed at keeping the dynamics as close as possible to the four-dimensional Robinson-Trautman equation. Similar to the four-dimensional case \cite{MacedoSaa}, equation (\ref{rtf}) represents the simplest local fourth-order scalar flow constructed from $P$ and the one-dimensional Laplacian $\Delta$ and which allows for the relevant stationary solutions, as we will demonstrate below. Furthermore, we will also show that the linearization of equation (\ref{rtf}) reduces to the Cahn-Hilliard equation, also mirroring the four-dimensional scenario.

 The first condition is
a direct consequence of the one-dimensional Stokes theorem applied to (\ref{ell1}) and (\ref{ell2}). As to the second condition, let us first look for the stationary solutions of (\ref{rtf}), which  can be written more explicitly  in the form
\begin{equation}
\label{rtf1}
 P_u = -\kappa P^2\partial_\phi P \partial_\phi \left( P + \partial_\phi^2 P \right).
\end{equation}
Its stationary solutions $P=P(\phi)$ must obey $  P \partial_\phi \left( P + \partial_\phi^2 P \right) = C$ constant, and consequently 
\begin{equation}
\label{C}
\partial_\phi \left( P + \partial_\phi^2 P \right)  = \frac{C}{P}.
\end{equation}
However, integrating both sides on the circle, we have that the left-hand side vanishes, while the left-hand side  will be $2\pi C$ assuming the normalization $\ell = 2\pi$, demanding $C=0$. Hence, we are left with $ P + \partial_\phi^2 P  = D$ constant, which has general periodic solution 
$P(\phi) = a + b\cos \phi + c\sin \phi$, with  constant $a$, $b$, and $c$. Imposing the length condition and redefining $\phi$, we obtain finally
(\ref{rocket}). The stationary solutions are thus uniquely the
normalized boosted BTZ configurations.

The full nonlinear stability analysis of (\ref{rtf1}) is highly non-trivial, but a useful first step is the linearized stability of the stationary family. Writing 
$P(u,\phi) = P_0(\phi) + \epsilon R(u,\phi)$
and retaining terms only to first order in $\epsilon$, one obtains
 a variant of the  Cahn-Hilliard equation \cite{CH}
\begin{equation}
\label{CH}
 R_u = -  P_0(\phi)^2\partial_\phi P_0(\phi) \partial_\phi \left( R + \partial_\phi^2 R \right),
\end{equation}
where we set $\kappa=1$ without loss of generality since it corresponds to a simple rescaling in $u$. 
For the isotropic asymptotic state ($v=0$ , $P_0=1$)
the general solution of (\ref{CH}) can be written in terms of a 
  Fourier decomposition 
\begin{equation}
R(u,\phi) = \sum_k a_k e^{-k^2(k^2-1)u}e^{ik\phi},
\end{equation}
where $a_k$ is to be determined by the initial data $R(0,\phi)$. As one can see, all modes with $|k|>1$ are exponentially suppressed, and consequently $R(u,\phi)$ tends asymptotically to the stationary solution, whose boundedness is enough to
establish that 
the isotropic state is   orbitally stable, {\em i.e.}, it is linearly stable modulo the neutral directions $k=\pm 1$, which are also associated with the stationary family.

When $v\neq0$, the coefficients in the partial differential equation (\ref{CH}) become $\phi$-dependent and the spectral analysis is no longer diagonal in the Fourier space. For small $v$ one may treat the anisotropy perturbatively, and the stability pattern remains the same. A fully non-perturbative proof for arbitrary $v$ is more subtle. In analogy with the Cahn-Hilliard equation, one may consider the quadratic functional (see 
\cite{CH} for further details)
\begin{equation}
\mathcal{E} = \frac{1}{2}\int_0^{2\pi} \left( R_\phi^2 - R^2 \right) d\phi,
\end{equation}
which is not bounded from below, demanding coercivity arguments and  rendering the analysis much more involved. Anyway, we could solve the full nonlinear equation (\ref{rtf1}) numerically for many smooth initial conditions, and for all of them we have observed a smooth relaxation towards a member of the family   (\ref{rocket}).

{
Let us also discuss more explicitly the positivity assumption on $P$. 
Notice first that regular solutions $P(u,\phi)$ cannot change sign since 
  the length (\ref{ell}) is conserved. In order to show this, let us consider that $P(0,\phi)>0$ and suppose, by contradiction, that  
 the solution attains $P(u,\phi) = 0$ for the first time for some $u=u_0$ at $\phi=\phi_0$. Choose a sequence $u_n$ converging to
 $u_0$ and let
 \begin{equation}
m_n=\min_\phi P(u_n,\phi),
 \end{equation}
 with $\phi_n$ corresponding to this minimum. We have $m_n\to 0$ and $\phi_n\to \phi_0$. Since the solution is
 assumed to be regular, near a minimum point $\phi_n$, Taylor's theorem gives
\begin{equation}
P(u_n,\phi_n+x)\leq m_n+C x^2
\end{equation}
for some constant $C>0$. Therefore
\begin{equation}
\ell(0)=\ell(u_n)
\geq
\int_{-\delta}^{\delta}
\frac{dx}{m_n+C x^2},
\end{equation}
for some $\delta$, 
which diverges as $u_n\to u_0$ and consequently $m_n\to 0$, contradicting the conservation of $\ell(u)$. Thus strictly positive initial data remain strictly positive four as long as the smooth solution exists. This argument does not, of course, replace a global existence theorem for the fourth-order quasilinear equation. Indeed, a rigorous global existence and positivity theorem for arbitrary smooth data is beyond the scope of the present work.
}

{
\subsection{Algebraic classification}

It is useful to determine the algebraic type of the spacetime (\ref{metric}) in the Cotton-scalar classification of $(2+1)$-dimensional geometries \cite{CottonPRD,CottonCQG}. We use the natural null triad
\begin{equation}
\label{triad}
k=\partial_r,\qquad
l=\partial_u+\frac{1}{2}\left(m_0+2r(\ln P)_u+\Lambda r^2\right)\partial_r,\qquad
m=\frac{P}{r}\partial_\phi,
\end{equation}
which satisfies $k\cdot l=-1$ and $m\cdot m=1$. For the convention
\begin{equation}
C_{abc}=2\left(\nabla_{[a}R_{b]c}-\frac{1}{4}\nabla_{[a}R\,g_{b]c}\right),
\end{equation}
the non-vanishing coordinate components are
\begin{equation}
C_{uru}=\frac{8\pi\mathcal{N}}{r^2},\qquad
C_{u\phi u}=-\frac{8\pi\mathcal{N}_{,\phi}}{r},\qquad
C_{u\phi\phi}=\frac{8\pi\mathcal{N}}{P^2}.
\end{equation}
together with the components fixed by antisymmetry in the first two indices and the identities of the Cotton tensor. The corresponding Cotton scalars are
\begin{equation}
\label{cottonscalars}
\Psi_0=\Psi_1=\Psi_2=0,\qquad
\Psi_3=\frac{8\pi\mathcal{N}}{r^2},\qquad
\Psi_4=-\frac{8\pi P\mathcal{N}_{,\phi}}{r^2}.
\end{equation}
Consequently, a generic non-stationary member of the null-fluid family is algebraically special of type III, with $k=\partial_r$ as a triple Cotton-aligned null direction. At points where $\mathcal{N}=0$ but $\mathcal{N}_{,\phi}\ne0$, the type degenerates to N. If $\mathcal{N}=0=\mathcal{N}_{,\phi}$, the Cotton tensor vanishes and the geometry is of type O. In particular, all stationary configurations (\ref{rocket}) have $\mathcal{N}=0$ and are conformally flat, as expected for locally BTZ geometries.
}

\subsection{The numerical solutions}

\begin{figure}[ht]
\includegraphics[width=0.49\linewidth]{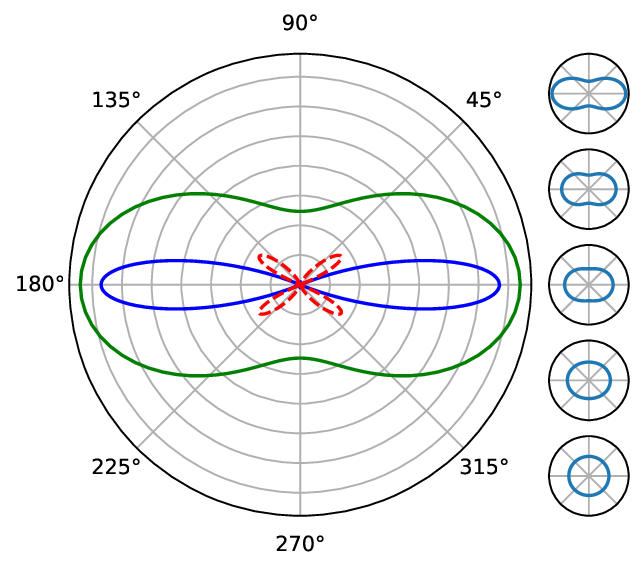}
\includegraphics[width=0.49\linewidth]{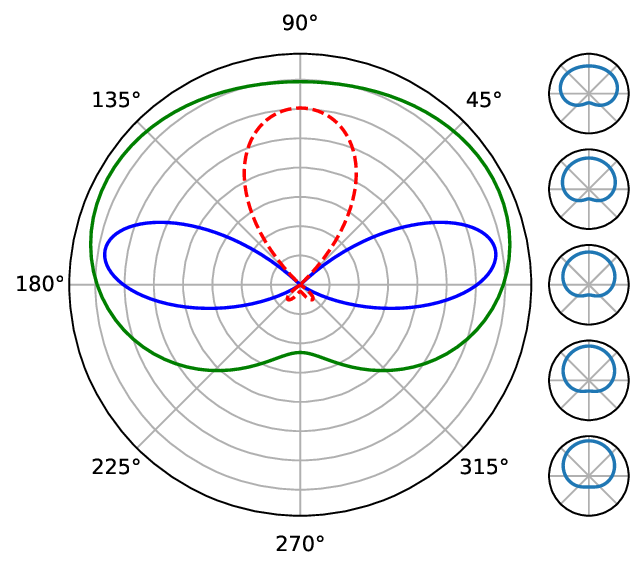} 
\caption{\label{fig1} Representative numerical evolutions of the flow (\ref{rtf1}). In the left panel the initial data
(\ref{P1}) relax toward the isotropic stationary state with $v=0$. In the right panel, on the other hand, the initial condition
(\ref{P2})   relaxes toward an anisotropic stationary state  (\ref{rocket})  with $v\approx0.52$. The successive insets illustrate, from top to bottom, the smoothing of the angular profile towards a stationary solution as the flow approaches its late $u$ limit. 
In both panels, the green (external) curves correspond to the respective initial conditions, while the internal curves
correspond to the right-hand side of the flow (\ref{rtf1}) at $u=0$, without scale. Blue (continuous) lines are the negative values, which makes $P(u,\phi)$ decrease  locally, while the red (dashed) are the positive values, associated with the local increase of
$P(u,\phi)$.  For an animation and further computational details, see \cite{RT3Site}.
}
\end{figure}

To illustrate the nonlinear dynamics we integrated equation (\ref{rtf1}) numerically for smooth periodic initial data. The angular derivatives were discretized with centered finite differences on a uniform grid over the circle, while for the 
$u$-evolution   we employed  a simple explicit forward step. For fourth-order parabolic problems, such explicit schemes are usually subject to a severe stability restriction of the form $\delta u\lesssim c(\delta\phi)^4$ (see, for instance, \cite{Num}). In our computations we obtained stable and accurate evolutions with $c\approx10^{-1}$, in agreement with the standard expectation for this class of equations. It is important to notice that 
the purpose of our numerical computations was  to show the qualitative relaxation mechanism, not to provide a definitive numerical study of the nonlinear fourth-order parabolic equation (\ref{rtf1}),
which is indeed a much more involved problem. We performed exhaustive  tests and
obtained a rather  robust pattern: smooth initial data do evolve smoothly  and approach the stationary family identified analytically above.
Fig. \ref{fig1} 
illustrates the flow evolution for some typical cases. The considered cases correspond to the initial data
\begin{equation}
\label{P1}
P(0,\phi) = C(1 + \cos^2\phi + \cos^4\phi)
\end{equation}
for the left panel,  and
\begin{equation}
\label{P2}
P(0,\phi) = C(2 + \cos^2\phi + \sin\phi)
\end{equation}
for the right panel. In both cases, the constant $C$ is determined to assure the normalization
$\ell(u) = 2\pi$, which we indeed   used   to control the numerical accuracy of the solutions. In both cases the numerical evolution preserves regularity and converges toward the stationary set. Additional computational details and animations are available in  \cite{RT3Site}.
{
In particular, in order to emphasize that the relaxation is not tied to the two profiles displayed in Fig.~\ref{fig1}, we also present the following additional cases corresponding to the following smooth positive normalized initial data:
\begin{eqnarray}
\label{P3}
P(0,\phi)&=&C\exp\left[0.35\cos(3\phi)+0.20\sin(5\phi)\right],\\
\label{P4}
P(0,\phi)&=&C\left[1+0.25\cos(\phi-0.7)+0.15\cos(4\phi)\right],\\
\label{P5}
P(0,\phi)&=&C\left[1+0.20\cos\phi+0.15\sin(2\phi)+0.10\cos(5\phi)\right].
\end{eqnarray}
The corresponding evolutions are summarized in Fig.~\ref{fig2}. Table~\ref{tab:numerics}
exhibits some numerical details of the evolutions depicted here.   
In all 
our exhaustive simulations, $P$ remains positive, the length constraint is preserved to reasonable accuracies, and the solutions always  approached the stationary family (\ref{rocket}).
\begin{figure}[ht]
\includegraphics[width=0.33\linewidth]{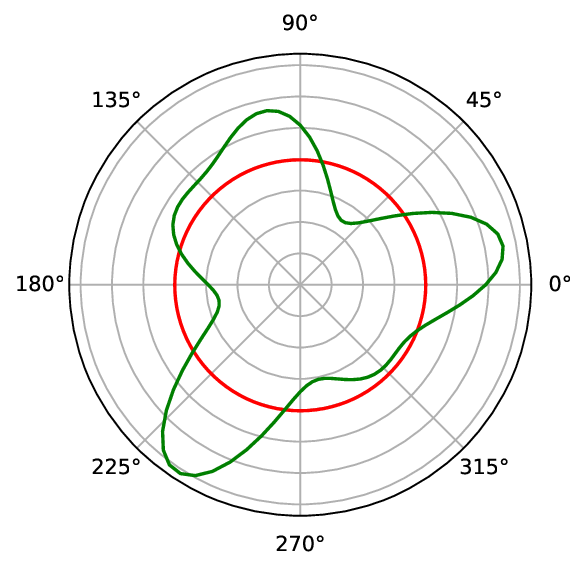}
\includegraphics[width=0.33\linewidth]{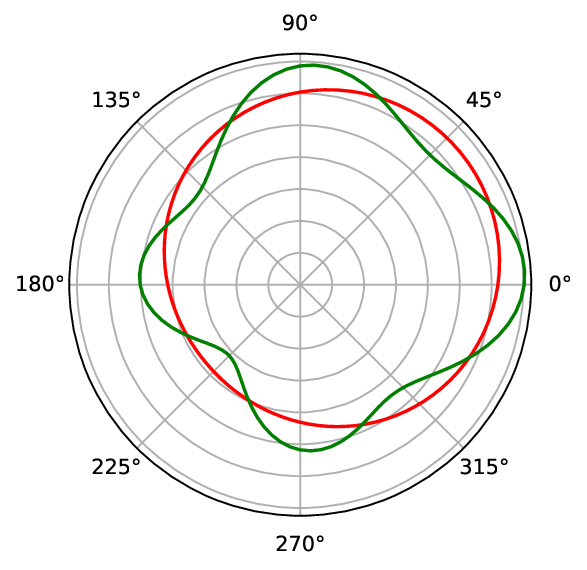} 
\includegraphics[width=0.33\linewidth]{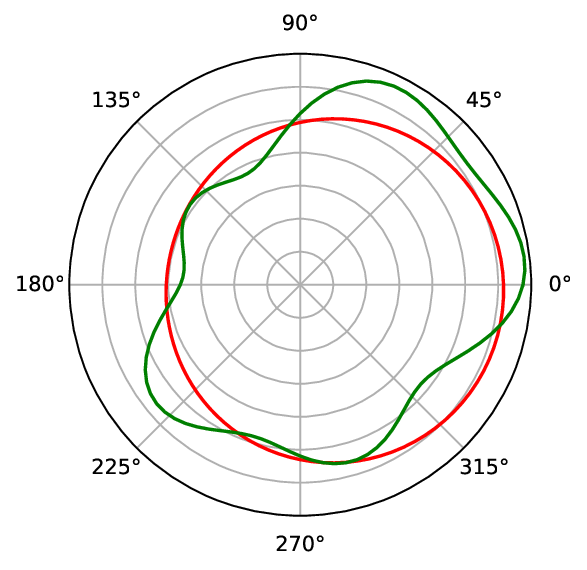} 
\caption{\label{fig2} Additional representative numerical evolutions of the flow (\ref{rtf1}) for the initial data (\ref{P3})--(\ref{P5}), corresponding, respectively, to the first, second, and third polar graphics. The solid curves show the initial data (green) and the late-time profile (red). The first profile, despite being non-symmetric (see the Final Remarks Section), relaxes to an almost isotropic state, while the second and third profiles relax to clearly boosted BTZ states.
These are some representative examples of our   exhaustive simulations, which exhibit a rather 
robust pattern: smooth initial data always  evolve smoothly and asymptotically approach the stationary family
(\ref{rocket}).
}
\end{figure}
\begin{table}[ht]
\caption{\label{tab:numerics} Some numerical data and diagnostics  for 
all evolutions considered here.  The values correspond to the final integration time $u_{\rm final}$
necessary to attain $\|P_u(u_{\rm final},\cdot)\|_2\le 10^{-4}$, with $\delta \phi = \pi/125$ and
$\delta u = (\delta \phi)^4/10$. For all cases,
we observed and error $\varepsilon = \left| \frac{\ell(u)}{2\pi} - 1 \right|\le 10^{-6}$ along  all the evolution.}
\begin{center}
\begin{tabular}{c|c|c|c|c}
case & $v $ & $\phi_{0  }$ & $u_{\rm final}$ & $ \varepsilon$  \\   \hline
(\ref{P1}) & $1.25\times10^{-14}$ & 3.04 & 0.970 & $6\times 10^{-8}$   \\ 
(\ref{P2}) & 0.522 & 1.57 ($\pi/2$) & 1.01 & $6\times 10^{-9}$  \\ 
(\ref{P3}) & $4.29\times 10^{-3}$& -1.58  & 0.766 & $4\times 10^{-7}$   \\ 
(\ref{P4}) & 0.258 & 0.703 & 0.694 & $5\times 10^{-7}$   \\ 
(\ref{P5}) & 0.209 & -0.183 & 0.861 & $7\times 10^{-7}$   \\ 
\end{tabular}
\end{center}
\end{table}
}

\section{Final remarks}

The proposed flow (\ref{rtf1}) defines a simple lower-dimensional Robinson-Trautman-like dynamics with a  clear physical interpretation. Its most interesting feature is 
a non-trivial dissipative dynamics which mimics
several qualitative features usually associated with gravitational radiation, which is not really available
in $(2+1)$ dimensions. 
 The dissipative dynamics is carried  by a null fluid, and the spacetime geometry relaxes toward a BTZ remnant, which
   may in general be moving with constant speed if the null fluid radiation is not symmetric. In this sense, the model offers a useful toy problem for studying how lower-dimensional gravity can encode anisotropic emission, symmetry constraints, and asymptotic state selection in a setting where the underlying PDEs are analogous to the real $(3+1)$ case. Several possible extensions suggest themselves. One may investigate more deeply  
the stability of the   stationary family 
by exploring entropy-like Lyapunov functionals  for the full nonlinear flow, and whether analogous flows 
may
arise in other matter-coupled lower-dimensional models. These questions would help clarify if  the present construction captures the essential part of Robinson-Trautman relaxation and to what extent it is special to the null-fluid realization made possible by the exact solutions of Ref.~\cite{2+1}.

 We wish to close with two final remarks. 
 The first   concerns the selection of the asymptotic state by symmetry arguments. 
Equation  (\ref{rtf1}) is invariant under the transformation  $\phi \to -\phi +\phi_1$,
which combines a rigid rotation with a reflection about the horizontal axis in the plane. Therefore, if the initial profile $P(0,\phi)$ is invariant under this transformation for some $\phi_1$, the entire evolution preserves that symmetry. The asymptotic state (\ref{rocket}) must then respect the very same reflection symmetry, implying that the remnant BTZ black hole can move only along the direction selected by $\phi_1$. If the initial data possess two or more distinct reflection symmetries, the only compatible late-time configuration is the isotropic one with $v=0$. This mirrors the role of discrete symmetries in the four-dimensional Robinson-Trautman recoil problem, see \cite{MacedoSaa}.
The second remark is the natural open question on whether the flow (\ref{rtf1}) would admit a direct interpretation as a geometric evolution of plane curves. If one interprets $\ell(u)$ as the arclength of a one-parameter family of planar curves $r_u(\phi)$ written in polar coordinates, then one would need
\begin{equation}
r_u(\phi)^2+\left(\frac{\mathrm{d}r_u}{\mathrm{d}\phi}\right)^2=\frac{1}{P(u,\phi)^2}.
\label{eq:curve}
\end{equation}
 We could not interpret the flow (\ref{rtf1})  solely in terms of the geometric invariants of the associated curves
$r_u(\phi)$. Establishing such a geometric reformulation would be very interesting, both because it might place the flow within the broader theory of curve evolution and because it could lead to sharper existence, uniqueness, and asymptotic results.

\section*{Acknowledgments}
A.S. is partially       supported by CNPq (Brazil) grant 306785/2022-6 
and wishes to thank Profs. Vitor Cardoso and Jose S. Lemos for the warm hospitality
at the Center for Astrophysics and Gravitation of the
University of Lisbon, where part of this work was done.

\end{document}